\documentclass[usenatbib]{mn}
\usepackage{psfig}

\footnotesize

\newcommand\nuddd{\ifmmode\stackrel{\bf \,...}{\textstyle \nu}\else$\stackrel{\,...}{\textstyle \nu}$\fi}
\def\lsim{~\rlap{$<$}{\lower 1.0ex\hbox{$\sim$}}}
\def\gsim{~\rlap{$>$}{\lower 1.0ex\hbox{$\sim$}}}

\title{Observations of Six Glitches in PSR B1737$-$30}

\author[Zou et al.]{W. Z. Zou,$^{1,4}$\thanks{Email: zouwz@uao.ac.cn}
~~N. Wang,$^{1}$~~R. N. Manchester,$^{2}$~~J. O. Urama,$^{1,3}$~~G.
Hobbs,$^{2}$
\newauthor Z. Y. Liu,$^{1}$ and J. P. Yuan$^{1}$\\
$^{1}$ Urumqi Observatory, NAO-CAS, 40-5 South Beijing
Road, Urumqi 830011, China\\
$^{2}$ Australia Telescope National Facility, CSIRO, PO Box 76,
Epping, NSW 1710, Australia\\
$^{3}$ Department of Physics \& Astronomy, University of Nigeria, Nsukka\\
$^{4}$ Graduate School of CAS, 19 Yuquan Road, Beijing, 100039,
China\\}

\begin{document}
\maketitle
\pagestyle{plain}

\begin{abstract}
Six glitches have been recently observed in the rotational frequency
of the young pulsar PSR B1737$-$30 (J1740$-$3015) using the 25-m
Nanshan telescope of Urumqi Observatory. With a total of 20 glitches
in 20\,years, it is one of the most frequently glitching pulsars of
the $\sim$1750 known pulsars. Glitch amplitudes are very variable
with fractional increases in rotation rate ranging from $10^{-9}$ to
$10^{-6}$. Inter-glitch intervals are also very variable, but no
relationship is observed between interval and the size of the
preceding glitch. There is a persistent increase in $|\dot\nu|$,
opposite in sign to that expected from slowdown with a positive
braking index, which may result from changes in the effective
magnetic dipole moment of the star during the glitch.

\end{abstract}


\section{Introduction}
Glitches in pulsars are rare and extraordinary events. The study of
such glitches and their post-glitch recoveries can give an insight
into the interior of neutron stars, the physics of ultradense matter
and provide limits on the equation of state
\citep{pin91,accp93,fle00}. Pulsar glitches are usually detected
from long-term and frequent timing observations. The number of
observed glitches has increased dramatically in the past few years,
which probably is due to both the increased continuous observations
of active pulsars and a large number of new young pulsars detected
by the pulsar surveys \citep[e.g., the Parkes surveys, see][]{kbm+03}
over the last decade.

Glitches are characterized as sudden increases in the rotation
frequency $\nu=1/P$ (where $P$ is the pulsar period), often followed
by an interval of approximately exponential recovery or relaxation
back towards the pre-glitch frequency. The post-glitch relaxation
can have timescale of days to years. Frequency jumps with magnitudes
$\Delta\nu/\nu \sim 10^{-6}$ are recognized as ``giant'' glitches,
and have been observed mostly in pulsars with characteristic ages
$\tau_c = P/(2\dot{P}) \sim 10^4$~yr, such as PSRs~B0833$-$45 (the
Vela pulsar), B1046$-$58 and B1338$-$62 \citep{wmp+00}. Giant
glitches have not been observed in the youngest radio pulsars, such
as PSR~B0531+21 (the Crab pulsar) and PSR~B1509$-$58. Detected
glitches are generally of magnitude $10^{-9}<\Delta\nu/\nu<10^{-6}$,
and the relative increment in slow-down rate $\Delta\dot\nu/\dot\nu$
is in the range $10^{-3}$ to $10^{-2}$. Very small glitches with
$\Delta\nu/\nu<10^{-9}$ are difficult to distinguish from timing
noise in young pulsars \citep{hlk06}. Of the 1750 known pulsars,
about 170 glitches in 54 pulsars have been observed so
far\footnote{See
  http://www.atnf.csiro.au/research/pulsar/psrcat/ and
  \citet{mhth05}}.  Half of the published glitches have fractional
amplitudes $\Delta\nu/\nu\sim 10^{-8}$. Most glitches detected in the
Vela pulsar are giant glitches, and the largest glitch, with
$\Delta\nu/\nu\approx16\times10^{-6}$, was observed in PSR
J1806$-$2125 \citep{hlj+02}. About 60\% of the 54 glitching pulsars
have glitched only once, and 16\% have glitched twice.  Observations
with Rossi X-ray Timing Explorer (RXTE) have revealed that PSR
J0537$-$6910, the 16-ms pulsar associated with the supernova remnant
N157B in the Large Magellanic Cloud, is the most frequently glitching
known pulsar with 23 glitches detected in seven years of monitoring
\citep{mmw+06}. Most of the glitches have $\Delta\nu/\nu$ of a
few$\times 10^{-7}$ and, interestingly, a strong correlation is
observed between the amplitude of a glitch and the time to the next
one. Furthermore, despite the usual post-glitch recovery (decrease) in
$|\dot\nu|$, a persistent long-term increase in this parameter is
observed, corresponding to a negative braking index
$n=\nu\ddot\nu/{\dot\nu}^2 \sim -1.5$. This increase is most probably
related to changes on or in the neutron star that result from the glitch
activity.

\begin{figure*}
\centerline{\psfig{file=glt1.ps,width=60mm,angle=-90}
\psfig{file=glt2.ps,width=60mm,angle=-90}
\psfig{file=glt3.ps,width=60mm,angle=-90}}
\end{figure*}
\begin{figure*}
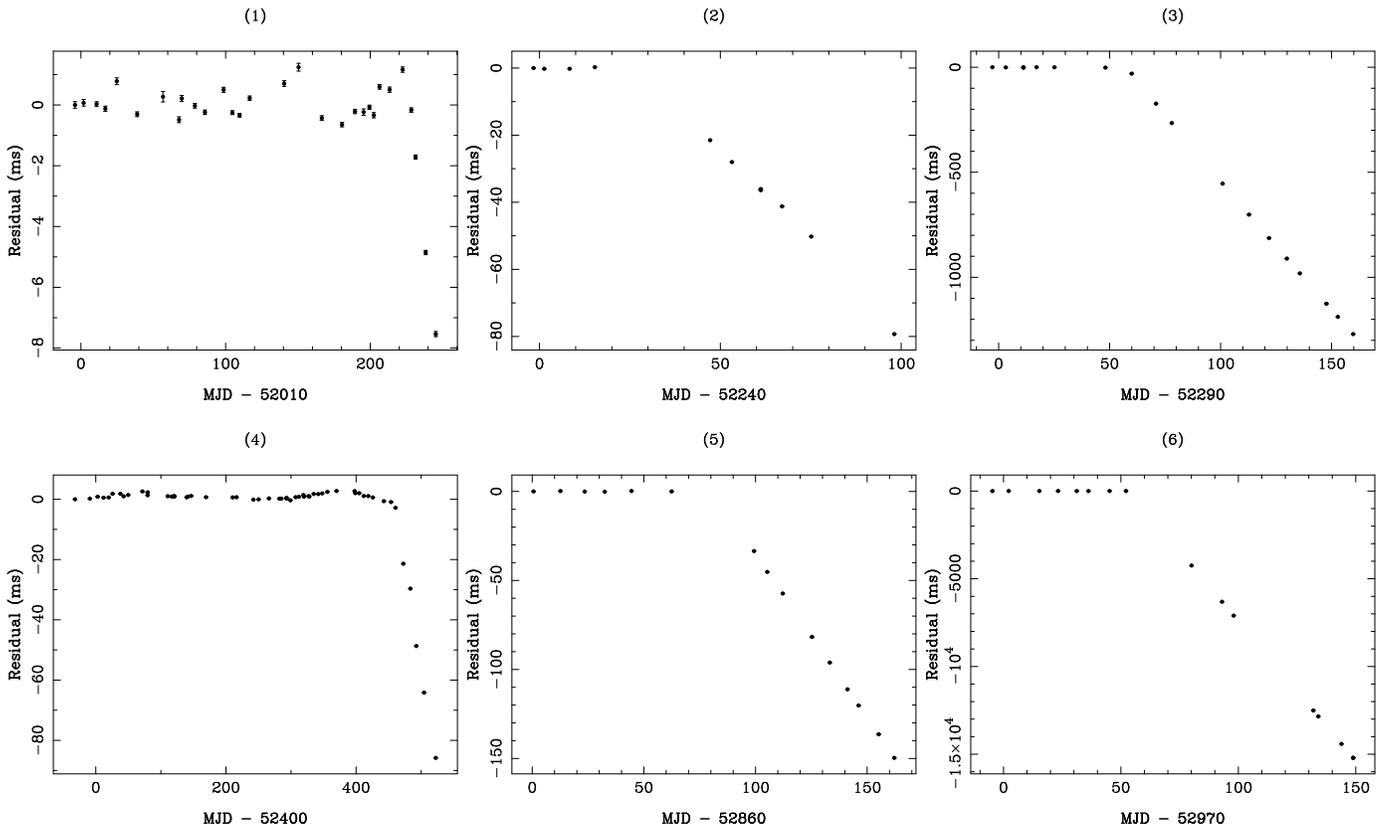

\centerline{\psfig{file=glt4.ps,width=60mm,angle=-90}
\psfig{file=glt5.ps,width=60mm,angle=-90}
\psfig{file=glt6.ps,width=60mm,angle=-90}} \caption{Timing residuals
relative to the pre-glitch timing model for
  six glitches in the rotation history of PSR B1737$-$30. Residuals
  are defined in the sense of observed arrival time minus predicted
  arrival time. }\label{fg:gl_phs}
\end{figure*}

In this paper we report on six recent glitches detected in the period
of PSR~B1737$-$30 and discuss the implications of these
results. PSR~B1737$-$30 is a young radio pulsar with a characteristic
age of $2.06\times10^{4}$~yr which glitches frequently, with 20 glitches of all sizes ranging from ``small'' to ``giant''
($\Delta\nu/\nu \sim 10^{-9}$ to $10^{-6}$) observed since its
discovery in 1986 \citep[][this paper]{ml90,sl96,klgj03}. As for PSR J0537$-$6910,
there appears to be a cumulative shift in the spin-down rate of PSR
B1737$-$30 resulting from its frequent glitches, which most likely
accounts for its long-term braking index of $n = -4\pm2$
\citep{ura02}.

\section{Observations and Analysis}

The timing observations of PSR B1737$-$30 using the 25-m Nanshan
radio telescope of Urumqi Observatory started in 2000 January using
a room-temperature dual-channel receiver. From 2002 July a
dual-channel cryogenic receiver system sensitive to two orthogonal
polarizations was used at a central observing frequency of 1540 MHz.
Two polarizations, each of bandwidth 320 MHz, are fed to a
filterbank consisting of $2\times128$ channels of width 2.5 MHz. The
data are digitized to one-bit precision with a sampling interval of
1 ms. Time is provided by a hydrogen maser clock calibrated using
the Global Positioning System. Observations times for PSR B1737-30
were generally 16 min with the room-temperature receiver before 2002
July and 4 min with the cryogenic receiver after that date.

The data are dedispersed with off-line programs to remove the
effects of interstellar dispersion, and then folded at the pulsar's
nominal topocentric period with four sub-integrations per
observation. The pulse profiles obtained by summing an observation
were cross-correlated with a high signal-to-noise template of the
pulsar profile to produce accurate pulse topocentric times of
arrival (TOAs), which are then processed with the standard timing
program TEMPO\footnote{see
http://www.atnf.csiro.au/research/pulsar/tempo/} to convert them to
barycentric arrival times at infinite frequency. The TOAs refer to
the peak of the main pulse. The Jet Propulsion Laboratory ephemeris
DE405 is used to correct the TOAs to the Solar system barycenter.
The folded profiles are stored on disk for subsequent processing and
analysis. Timing observations are usually made approximately three
times per month and the glitching pulsars are observed more
frequently.

We use the corrected barycentric TOAs to determine the basic
parameters of the pulsar. The basic timing model gives the predicted
rotational pulse phase $\phi_m(t)$, as a function of time, t:
\begin{equation}
\phi_m(t) = \phi_0 + \nu_0(t-t_0) + \frac{1}{2}\dot\nu_0(t-t_0)^2
+ \frac{1}{6}\ddot\nu_0(t-t_0)^3.
 \label{eq:model}
\end{equation}
Timing irregularities appear as phase residuals ($\phi-\phi_m$), which
are usually divided by $\nu$ to place them in time units.

For a glitching pulsar the residuals suddenly develop a negative slope
at the time of the glitch as illustrated in Fig.~\ref{fg:gl_phs}. The
frequency perturbation due to the glitch can usually be described by:
\begin{equation}
\Delta\nu(t)
=\Delta\nu_p + \Delta\dot\nu_p\,t + \Delta\nu_d\exp(-t/\tau_d),
 \label{eq:gl_model}
\end{equation}
where $\Delta\nu =\nu-\nu_0$ is the change in pulse frequency relative
to the pre-glitch model, $\Delta\nu_p$ and $\Delta\dot\nu_p$ are the
permanent changes in frequency and frequency derivative respectively
and $\Delta\nu_d$ is the amplitude of the exponential recovery with
a decay time constant of $\tau_d$. The total frequency jump at the
time of the glitch $\Delta\nu_g = \Delta\nu_p + \Delta\nu_d$ and the
degree of recovery is often described by the parameter
$Q=\Delta\nu_d /\Delta\nu_g$. Because of the decaying component, the
instantaneous change in $\dot\nu$ at the glitch differs from
$\Delta\dot\nu_p$:
\begin{equation}
\Delta\dot\nu_g = \Delta\dot\nu_p - Q\Delta\nu_g/\tau_d.
 \label{eq:dnudot}
\end{equation}
The glitch model of Equation \ref{eq:gl_model}
describes the post-glitch behaviour fairly well for most glitches
\citep{sl96,wmp+00}.

\begin{table*}
\begin{minipage}{13cm}
\caption{The rotation parameters for PSR B1737$-$30. The errors
are at 2$\sigma$ level.}
\begin{tabular}{clllcccc}
\hline
Fit Span     & Epoch   & \multicolumn{1}{c}{$\nu$}        & \multicolumn{1}{c}{$\dot\nu$}            & \multicolumn{1}{c}{$\ddot\nu$}              & Residual  & No. of \\
(MJD)        & (MJD)   & \multicolumn{1}{c}{($s^{-1}$)}   & \multicolumn{1}{c}{($10^{-12}$s$^{-2}$)} & \multicolumn{1}{c}{($10^{-24}$s$^{-3}$)}    & ($\mu$s)      & TOAs\\
\hline
51549-52233  & 51891   & 1.64806684404(3)                 & $-$1.266582(5)                             &  0.6(3)                                     & 1012          & 79 \\
52238-52256  & 52247   & 1.6480278954(3)                  & $-$1.2688(7)                               &   --                                        & 97            & 4  \\
52287-52339  & 52313   & 1.6480206942(2)                  & $-$1.2650(3)                               &   --                                        & 369           & 8  \\
52349-52854  & 52602   & 1.64798930701(5)                 & $-$1.266509(4)                             &  47(1)                                      & 1223          & 51 \\
52860-52923  & 52892   & 1.6479576154(1)                  & $-$1.2666(1)                               &   --                                        & 137           & 6  \\
52959-53023  & 52991   & 1.64794681727(8)                 & $-$1.26667(9)                              &   --                                        & 147           & 8  \\
53050-54094  & 53572   & 1.64788626572(4)                 & $-$1.266449(2)                             &  15.1(2)                                    & 2722          & 103 \\
\hline
\end{tabular}
\label{tb:rotn}
\end{minipage}
\end{table*}

\begin{table*}
\begin{minipage}{14.5cm}
\caption{The glitch parameters of PSR B1737$-$30. The errors
are at 2$\sigma$ level. }
\begin{tabular}{lllllrccrc}
\hline
\multicolumn{1}{c}{Glitch}  &
\multicolumn{1}{c}{Glitch epoch}  & \multicolumn{1}{c}{Date}     & \multicolumn{1}{c}{Fit Span}        & \multicolumn{1}{c}{$\Delta\nu_g/\nu$}   & \multicolumn{1}{c}{$\Delta\dot\nu_p/\dot\nu$}   & \multicolumn{1}{c}{$\tau_d$}  & Q  & \multicolumn{1}{c}{$\Delta\dot\nu_g/\dot\nu$}  & Residual  \\
  No.             & \multicolumn{1}{c}{(MJD)}  &           &
  \multicolumn{1}{c}{(MJD)}  & \multicolumn{1}{c}{($10^{-9}$)}
  & \multicolumn{1}{c}{($10^{-3}$)}                   &
  \multicolumn{1}{c}{(days)}      &       &
  \multicolumn{1}{c}{($10^{-3}$)}  & ($\mu$s) \\
\hline
(1)   &    52237(1)     & 011111   &  52005-52256  & 5.0(4)       & --
& --            & --   & \multicolumn{1}{c}{--}      & 353   \\
(2)   &    52260(4)     & 011217   &  52238-52339  & 12(4)        &
$-$3(3)        & --            & --   &  $-$3(3)       & 311   \\
(3)   &    52347.66(6)  & 020314   &  52287-52450  & 152(2)       &
$-$4.6(4)      & 50            & 0.103(9) & 0.1(7)  & 297   \\
(4)   &    52858(2)     & 030807   &  52367-52923  & 19(2)        &
1.0(6)         & --            & --   & 1.0(6)      & 1020  \\
(5)   &    52941.3(6)   & 031029   &  52860-53063  & 21.6(6)      & 0.4(2)         & --            & --   &  0.4(2)     & 219   \\
(6)   &    53036(13)    & 040201   &  52965-53218  & 1853.6(14)   &
$-$5.36(7)     & 100           & 0.0302(6) & 3.0(2) & 256  \\
\hline
\end{tabular}
\label{tb:glt}
\end{minipage}
\end{table*}

\section{Results}

Six glitches have been observed in the period of PSR B1737$-$30 over
the seven-year monitoring period from 2000 January 6 (MJD 51549) to
2006 December 25 (MJD 54094). The rotation history of these glitches
with respect to pre-glitch timing models are shown in
Fig.~\ref{fg:gl_phs}. No glitches have been observed since 2004
February, i.e., for nearly three years. Tables 1 and 2 give the main
parameters of the six glitches. Uncertainties are parentheses and
refer to the last quoted digit. Glitch parameters have been determined
by fitting Equation~\ref{eq:gl_model} to the timing data around each
glitch. For the smaller glitches, the glitch epoch is determined
unambiguously by the requirement of pulse phase continuity. The last
large glitch, the epoch is ambiguous and is given as the mean of the
dates of the last pre-glitch and the first post-glitch
observations. Except for the largest glitch, the magnitudes of these
glitches are comparable to those observed from Crab pulsar
($\sim10^{-8}$), but generally $2-3$ orders smaller than the typical
glitch of the Vela pulsar ($\Delta\nu/\nu\sim10^{-6}$). No changes
associated with the glitches in the pulse shape or flux density have
been observed.

\begin{figure}
\centerline{\psfig{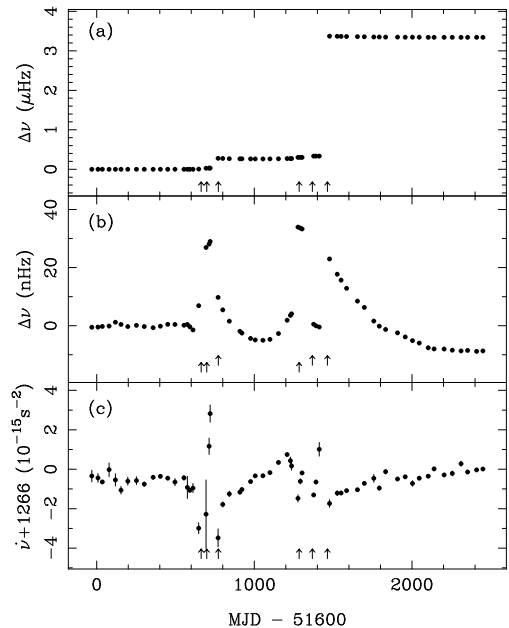}}
\caption{Pulse frequency variations for PSR B1737$-$30 over an
2500-d
  period: (a) frequency residual $\Delta\nu$ relative to the
  pre-glitch solution, (b) an expanded plot of $\Delta\nu$ where the
  mean residual between glitches with a raised arrow and the following
  glitch has been removed from data after the marked glitch, and (c)
  the variations of $\dot\nu$.}
\label{fg:freq_res}
\end{figure}

Fig.~\ref{fg:freq_res} shows the time-evolution of the residual
frequency ($\Delta\nu$) and the frequency first derivative
($\dot\nu$) over a 2500-d period. The six glitches are clustered in
two groups of three. The first observed glitch, at epoch MJD 52237
(2001 November 11, glitch 1 in Fig.~\ref{fg:gl_phs}) is very small,
with a fractional change in the rotation rate $\Delta\nu_g/\nu \sim
5\times10^{-9}$. A relatively large glitch ($\Delta\nu_g/\nu \sim
1.52\times10^{-7}$) occurred at epoch MJD 52347.66 (2002 March 14,
glitch 3 in Fig.~1), the amplitude of which is similar to a typical
glitch of PSR B1758$-$23; PSRs B1737$-$30 and B1758$-$23 have
similar properties --- age, period, period derivative and spin-down
energy loss rate. The relaxation after this glitch is adequately
modelled by an exponential decay of time scale $\sim50$~d with
$Q\sim 0.1$. The largest glitch so far observed in this pulsar with
$\Delta\nu/\nu = 1.85\times10^{-6}$ occurred on MJD $53036\pm13$
(2004 February 1, glitch 6 in Fig.~\ref{fg:gl_phs}). This is a
Vela-type glitch. Fitting glitch parameters to a relatively short
(200 d) data span gives a good fit to the data. However, the derived
value of $\Delta\dot\nu_p$ is highly covariant with the exponential
decay term; for the solution given in Table~\ref{tb:glt} we have
fixed the decay timescale at 100~d.

The two largest glitches observed are the third glitch (MJD
52347.66) and the final glitch (MJD 53036). Both are the final
glitch of their group. Following the short-term (50 -- 100~d)
exponential recovery, Fig.~\ref{fg:freq_res} shows that, for these
two glitches, there is an approximately linear increase in $\dot\nu$
until the next glitch. The slopes of these two linear regions are
not the same, with the earlier one having a gradient about four
times as large, despite it being the smaller glitch.

As we were completing this work, \citet{js06} published glitch
parameters on seven pulsars that include PSR B1737$-$30. Within the
data span covered by the Nanshan observations they report a total of
10 glitches. The first three of those glitches (between MJD 51685
and 52007) have $\Delta\nu/\nu < 10^{-9}$. Our data do not show any
significant jumps (above the level of timing noise) in this
interval. Such glitch magnitudes imply residual slopes of $\sim$
1\,ms in 20\,days. Our fit to the 700-d data span (MJD 51549 ---
52232; Table~\ref{tb:rotn}) has an rms residual of $\sim 1$~ms and
there is no evidence for discontinuities in the pulse phase. The
paper also reports two other small glitches, with $\Delta\nu/\nu
\sim 1.5\times10^{-9}$ around MJD 52603 and 52759. Similarly, we
find no evidence for discontinuities at these times (see residual
plot for glitch 4 in Fig.~\ref{fg:gl_phs}, Fig.~\ref{fg:freq_res}
and Table~\ref{tb:rotn}).  For three of the five significant
glitches (with $\Delta\nu/\nu > 10^{-9}$) in \citet{js06}, we obtain
glitch magnitudes that disagree substantially with the values given
in that paper, ours being generally lower. In two of the three cases
(glitches 1 and 2 in this paper), they differ by an order of
magnitude or more, while the glitch epochs are in good agreement.
For glitches 4 and 5 of this paper, the glitch magnitudes and epochs
are in good agreement with the \citet{js06} values. Our glitch 6 is
outside their data span. The differing results are probably a result
of sparse sampling in their early data (G. Janssen, private
communication).

\section{Discussion}
\begin{figure}
\centerline{\psfig{file=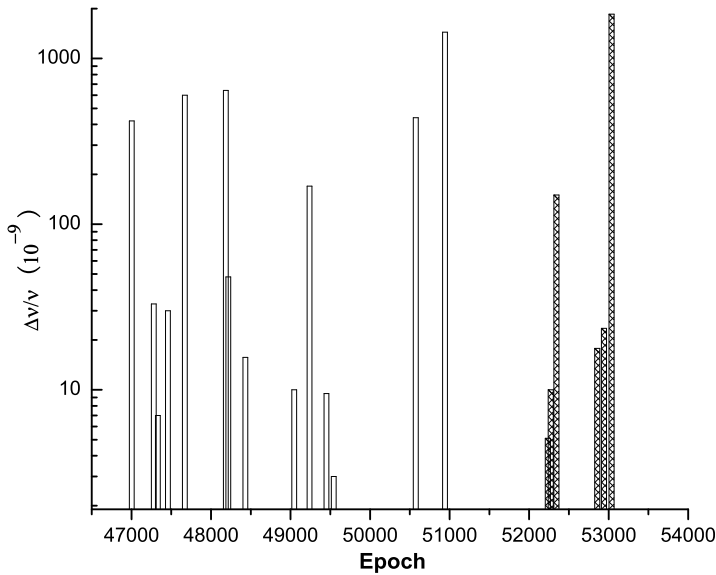,width=95mm,angle=0}}
\caption{The glitch history of PSR B1737$-$30. Glitches from MJD
52000 are from our work.} \label{fg:glitches}
\end{figure}

PSR B1737$-$30 is one of the most frequently glitching pulsars, and
it has a large range of glitch amplitudes. Table 3 presents the 14
glitches of PSR B1737$-$30 previously published by \citet{ml90},
\citet{sl96} and \citet{klgj03}. Adding the six glitches given in
Table~\ref{tb:prev_gl} brings the total to 20 glitches in the 20
years 1986 to 2006. The glitch history of $\Delta\nu/\nu$ for the 20
glitches is shown in Fig.~\ref{fg:glitches}. Unlike PSR J0537$-$6910
\citep{mmw+06}, there is no clear relationship of interval to the
next glitch to glitch size. In PSR B1737$-$30, the glitches tend to
cluster in groups. Sometimes the last glitch of a group is small
(e.g., the group between MJD 49000 and 49600) and sometimes it is
large (e.g., glitch 6 at MJD 53035).

For the Vela pulsar and some others, e.g. PSR B1800$-$21 (see the
table of glitches in the ATNF Pulsar Catalogue), the glitches are
either large ($\Delta\nu_g/\nu \gsim \; 10^{-6}$) or small ($\Delta\nu_g/\nu
\lsim \; 10^{-8}$). PSR J0537$-$6910 is similar except that the typical
large glitch is smaller, $\Delta\nu_g/\nu \sim 3\times 10^{-7}$. However, in common
with a number of other pulsars, e.g., PSRs B1046$-$58 and PSR
B1338$-$62, PSR B1737$-$30 has a more uniform distribution of glitch
sizes. Excepting PSR J0537$-$6910 which is younger, all of these pulsars have
characteristic ages in the range $1 - 2 \times 10^4$ years, so the
apparent difference in glitch properties is not simply a function of
(characteristic) age. The difference appears quite significant, but
its origin is unclear.

Derived values of $\Delta\dot\nu_g/\dot\nu$ are much less reliable
since they depend heavily on the post-glitch sampling and the
adequacy of the model for the post-glitch decay. Earlier values are
quite uncertain, but for most of the larger glitches
$\Delta\dot\nu_g/\dot\nu \sim 10^{-3}$, about an order of magnitude
less than the corresponding values for the Vela pulsar. The linear
increases in $\dot\nu$ in the intervals following the two large
glitches at MJD 52347 and 53036 are qualitatively similar to those
observed in the Vela pulsar \citep{lpgc96}. However, the linear
gradients observed in Vela are much steeper, typically $\sim 25
\times 10^{-15}$~s$^{-2}$~yr$^{-1}$ whereas, for PSR B1737$-$30,
even the steeper gradient following the glitch of MJD 52347 is just
$\sim 2\times 10^{-15}$~s$^{-2}$~yr$^{-1}$.

\begin{table}
\caption{Published glitches of PSR B1737$-$30.}\label{tb:prev_gl}
\begin{tabular}{lrrc}
\hline
Epoch          & \multicolumn{1}{c}{$\Delta\nu_g/\nu$}  & \multicolumn{1}{c}{$\Delta\dot\nu_g/\dot{\nu}$}   & Refs \\
MJD            & \multicolumn{1}{c}{($10^{-9}$)}        & \multicolumn{1}{c}{($10^{-3}$)}                   &      \\
\hline
47003(25)      & 420(20)                                & 3(1)                                              & 1    \\
47281(2)       & 33(5)                                  & 2(4)                                              & 1    \\
47332(16)      & 7(5)                                   & $-$1(12)                                          & 1    \\
47458(2)       & 30(8)                                  & 0(4)                                              & 1    \\
47670.2(2)     & 600.9(6)                               & 2.0(4)                                            & 1    \\
48186(6)       & 642(16)                                & $-$5(12)                                          & 2    \\
48218(2)       & 48(10)                                 & 8(12)                                             & 2    \\
48431(1)       & 15.7(5)                                & 0.8(3)                                            & 2    \\
49046(4)       & 10.0(4)                                & 0.01(6)                                           & 2    \\
49239(2)       & 169.6(3)                               & 0.8(1)                                            & 2    \\
49451.7(4)     & 9.5(5)                                 & $-$0.32(2)                                        & 3    \\
49543.93(8)    & 3.0(6)                                 & $-$0.68(2)                                        & 3    \\
50574.5497(4)  & 439.3(2)                               & 1.261(2)                                          & 3    \\
50941.6182(2)  & 1443.0(3)                              & 1.231(5)                                          & 3    \\
\hline
\end{tabular}
\center {References: 1. McKenna \& Lyne (1990); 2. Shemar \& Lyne
(1996); 3. Krawczyk et al. (2003).}
\end{table}
\nocite{ml90,sl96,klgj03}

The glitch activity, $A_g$, defined as the mean fractional change in
period per year owing to glitches \citep{ml90}, is given by the simple
expression:
\begin{equation}
A_g=\frac{1}{t_g}\sum{\frac{\Delta\nu_g}{\nu}},
 \label{eq:a_g}
\end{equation}
where $\sum\Delta\nu_g/\nu$ is the total fractional increase of
frequency owing to all of the glitches over an interval of $t_g$.
PSR B1737$-$30 has a relatively high glitch activity with
$A_g=2.96\times10^{-7}$ yr$^{-1}$. An advantage of $A_g$ as a
long-term indicator of glitch effects is that it is relatively
insensitive to the additional discovery of smaller glitches as the
data quality improves \citep{wbl01}.  Pulsars can be grouped into
three classes: pulsars with low glitch activity (e.g., PSR
B0525+21), high glitch activity (e.g., Vela pulsar) and no glitch
activity.  PSR B1737$-$30 belongs among those with high glitch
activity.  The pulsars with characteristic ages between $10^4$ and
$10^5$ yr are more likely to have higher glitch activity than the
younger ones ($\tau_c < 10^4$\,yr) or the older ones ($\tau_c >
10^5$\,yr) as shown by statistical studies of pulsar glitches
\citep{uo99,lsg00,wmp+00}.

\begin{figure}
\centerline{\psfig{file=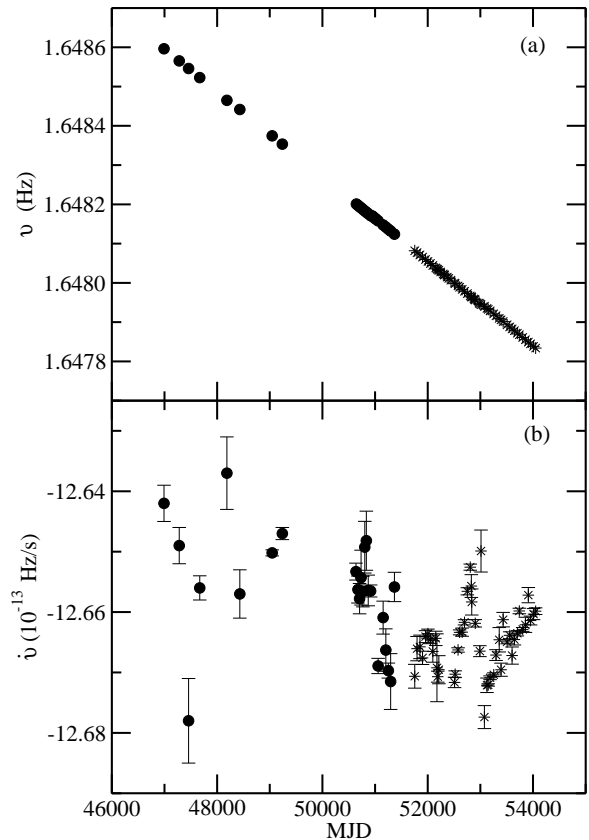,width=90mm,angle=0}}
\caption{Evolution of the spin parameters over 20 years (a) rotation
frequency $\nu$, and
  (b) spin-down rate, $\dot{\nu}$.  Points marked with a $\bullet$ are
  from Shemar \& Lyne (1996); Urama (2002) and Krawczyk et al. (2003);
  those with a $\star$ from the present work. We have omitted the
  measured $\dot{\nu}$ values $<$ 150 days after a glitch. }
      \label{fg:spin-rate}
\end{figure}

Fig.~\ref{fg:spin-rate} shows the evolution of the spin frequency
$\nu$ and the slow-down rate $\dot\nu$ over 20 years. Clearly, the
long-term evolution of $\nu$ is dominated by the regular spin-down --
the glitches are not even visible on this plot. The lower part of
Fig.~\ref{fg:spin-rate} shows that there has been a long-term decrease
in $\dot\nu$ (increase in $|\dot\nu|$) of about $\sim$ 0.1\% over the
20 years. This corresponds to a value of $\ddot\nu \sim
-3\times10^{-24}s^{-3}$. The braking index is therefore $-3\pm1$ for
the long-term evolution of PSR B1737$-$30.  Inter-glitch data spans
are frequently affected by relaxation from large glitches and
generally have large positive braking indices \citep{jg99,wmz+01}. For
example, we obtain a braking index of $n=13\pm1$ by fitting the data
starting 500 days after the last glitch (MJD 53550 -- 54095).

As mentioned in the Introduction, a negative long-term braking index
is also observed for PSR J0537$-$6910 \citep{mmw+06}. Permanent
increases in $|\dot\nu|$ associated with glitches are also observed
in the Crab pulsar \citep{lsg00,wbl01}. In this case though, since
the glitches are much smaller, they do not bias the observed braking
index so much. The Vela pulsar also has an anomalously low value of
$n$ \citep{lpgc96} which may be likewise related to the frequent
glitches. A possible interpretation of these observations is that
the component of the magnetic dipole moment which is perpendicular
to the spin axis increases at the time of a glitch
\citep{lfe98,rzc98}.

Fig.~\ref{fg:spin-rate}(b) also shows what appears to be a cyclic
variation in $\dot{\nu}$ superimposed over the linear trend. Such a
cyclic variation would have $\sim$ 3000\,d period and needs at least
another cycle to confirm.

\begin{figure*}
\centerline{\psfig{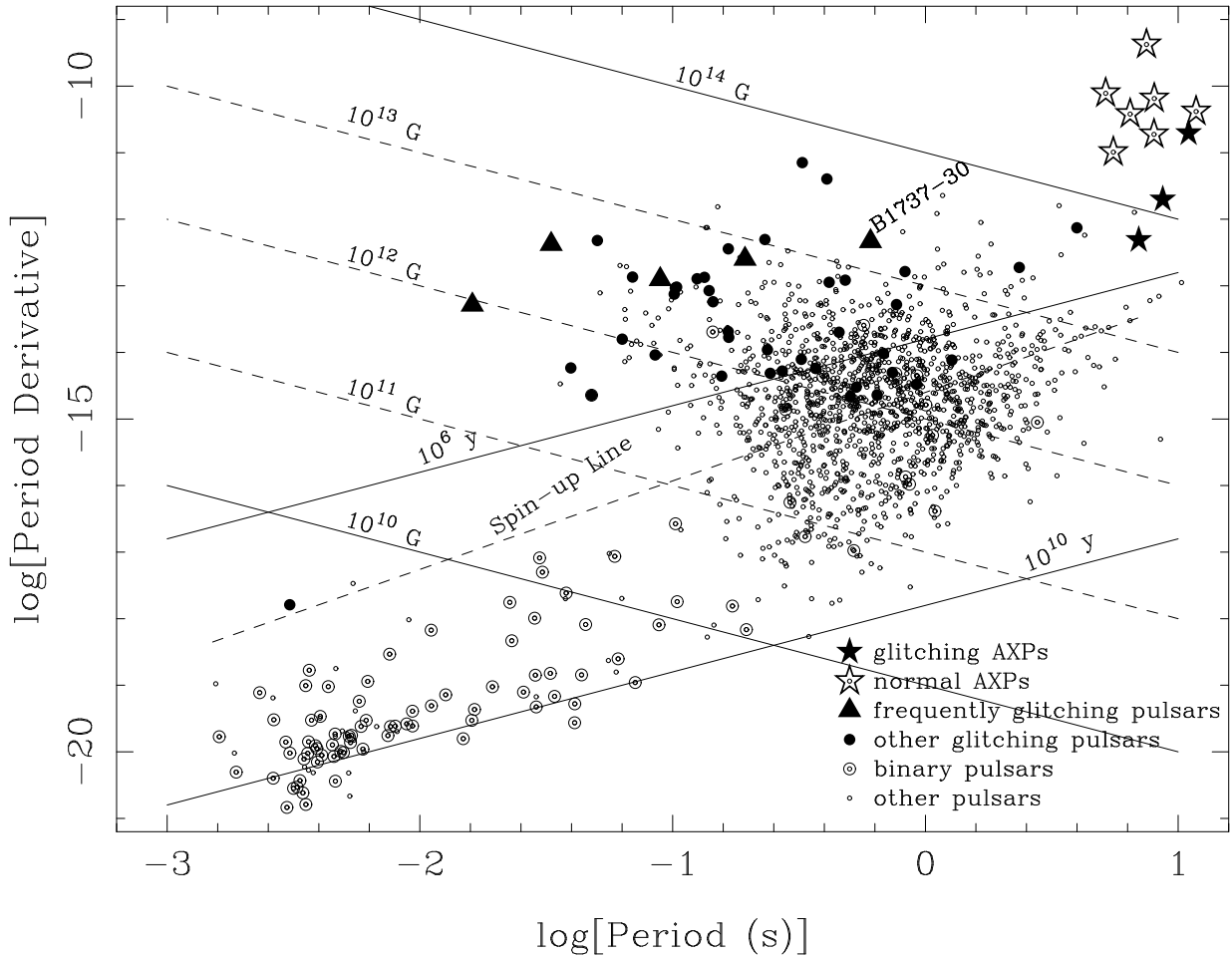}}
\caption{Distribution of known pulsars in the period ---
period-derivative plane. Glitching AXPs are shown by filled stars,
frequently glitching pulsars are indicated by triangles and the rest
of the  glitching pulsars are marked by bold points. Lines of
constant characteristic age and surface dipole magnetic field
strength are shown.} \label{fg:PPdot}
\end{figure*}

Fig. \ref{fg:PPdot} is a plot of pulsar period P versus period
derivative $\dot{P}$ of all known 1700 pulsars which distinguishes
the glitching pulsars and other different classes of pulsars. The
most frequently glitching pulsars are young and have relatively
strong dipole fields. Some glitching pulsars are older, including
PSR B1821$-$24, which is a recycled pulsar in the globular cluster
M28 \citep{cb04}. At present, 20 pulsars have detected giant
glitches and, except PSR B2224+65 whose age is $1.12\times10^6$ yr, all
the pulsars which have giant glitches are in the younger group.

\section{Conclusions}
Our observations of PSR B1737$-$30 show six glitches in PSR
B1737$-$30 over a period of seven years. In total, 20 glitches have
been observed over a 20-year period.  Glitch sizes cover a wide
range from just a few parts in $10^9$ to large glitches comparable
to those seen in the Vela pulsar. The inter-glitch intervals are
highly variable, ranging from 3 weeks to more than 3 years. Unlike
in PSR J0537$-$6910, there is no clear relationship of glitch
interval with size of the preceding glitch. Although the age of PSR
B1737$-$30 is comparable to that of the Vela pulsar, their glitch
behaviours are different, with glitch sizes and intervals being much
more variable in PSR B1737$-$30. Exponential and quasi-linear
relaxations in $\dot\nu$ are observed in both pulsars, but with
quite different parameters. It is remarkable that the three most
highly glitching pulsars have such different glitch and post-glitch
properties, yet another example of the diversity and complexity of
magnetised neutron stars.

\section*{ACKNOWLEDGEMENTS}
We acknowledge the support from the NNSFC under the project
10673021; and the Key Directional Project program of CAS. WZ is very
grateful to J. Yang for his valuable suggestions that led to an
improvement on an earlier draft of this paper. JOU is grateful for
the CAS--TWAS Visiting Scholar Fellowship that enabled him visit the
Urumqi Observatory of NAOC.


\end{document}